\documentclass[runningheads]{llncs}

\usepackage[utf8]{inputenc}
\usepackage{url}
\usepackage{listings}
\usepackage{booktabs}
\usepackage{algorithm}
\usepackage{algorithmic}
\usepackage{subcaption}
\usepackage{array}
\usepackage{lscape}
\usepackage{multirow, makecell}
\usepackage{backnaur}
\usepackage{graphicx}
\usepackage{amsmath}
\usepackage{amssymb}
\usepackage{xcolor} 
\usepackage{titlesec}
\usepackage{rotating}
\usepackage[bookmarks=true]{hyperref}
\hypersetup{
pdfpagemode={UseOutlines},
bookmarksopen=true,
bookmarksopenlevel=1,
colorlinks=true,% Set to false to disable coloring links
citecolor=blue,% The color of citations
linkcolor=blue,% The color of references to document elements (sections, figures, etc)
urlcolor=blue,% The color of hyperlinks (URLs)
breaklinks=true,
}
\newcommand\blfootnote[1]{%
  \begingroup
  \renewcommand\thefootnote{}\footnote{#1}%
  \addtocounter{footnote}{-1}%
  \endgroup
}

\begin{document}
\pagenumbering{gobble}

\title{Leveraging Large Language Models for Semantic Query Processing in a Scholarly Knowledge Graph}

\author{Runsong Jia\orcidID{0009-0003-3986-2679}\and 
Bowen Zhang\orcidID{0000-0001-6045-8599}\and 
Sergio J. Rodríguez Méndez\orcidID{0000-0001-7203-8399}\and
Pouya G. Omran\orcidID{0000-0002-4473-3877}}

\institute{Australian National University, Canberra ACT 2601, Australia
\url{https://comp.anu.edu.au/}
}

\authorrunning{Jia et al.}
\titlerunning{Semantic Query Processing with LLMs and KGs}
\maketitle

\begin{abstract}
\vspace{-6mm}
The proposed research aims to develop an innovative semantic query processing system that enables users to obtain comprehensive information about research works produced by Computer Science (CS) researchers at the Australian National University (ANU). The system integrates Large Language Models (LLMs) with the ANU Scholarly Knowledge Graph (ASKG), a structured repository of all research-related artifacts produced at ANU in the CS field. Each artifact and its parts are represented as textual nodes stored in a Knowledge Graph (KG).

To address the limitations of traditional scholarly KG construction and utilization methods, which often fail to capture fine-grained details, we propose a novel framework that integrates the Deep Document Model (DDM) for comprehensive document representation and the KG-enhanced Query Processing (KGQP) for optimized complex query handling. DDM enables a fine-grained representation of the hierarchical structure and semantic relationships within academic papers, while KGQP leverages the KG structure to improve query accuracy and efficiency with LLMs.

By combining the ASKG with LLMs, our approach enhances knowledge utilization and natural language understanding capabilities. The proposed system employs an automatic LLM-SPARQL fusion to retrieve relevant facts and textual nodes from the ASKG. Initial experiments demonstrate that our framework is superior to baseline methods in terms of accuracy retrieval and query efficiency.

We showcase the practical application of our framework in academic research scenarios, highlighting its potential to revolutionize scholarly knowledge management and discovery. This work empowers researchers to acquire and utilize knowledge from documents more effectively and provides a foundation for developing precise and reliable interactions with LLMs.

\end{abstract}

\blfootnote{Copyright © 2024 for this paper by its authors. Use permitted under Creative Commons License Attribution 4.0 International (CC BY 4.0).}

\vspace{-11mm}

\section{Introduction}
In the era of exploding information, KGs (KGs) have emerged as a powerful tool for organizing, representing, and analyzing complex data. Particularly in the domain of academic research, KGs have shown immense potential in facilitating knowledge discovery, information retrieval, and scholarly communication \cite{abu2021domain}. By integrating and linking vast amounts of academic data, such as research papers, authors, institutions, and research topics, academic KGs provide a structured and machine-readable representation of scholarly knowledge \cite{ostendorff2019enriching}. Moreover, academic KGs offer unique value and application prospects in academic research, such as promoting cross-disciplinary knowledge discovery and supporting scientific research decisions.

However, despite the promising applications of KGs in academia, traditional KG construction and utilization methods face several limitations when dealing with academic literature. One major challenge lies in the granularity of information representation \cite{wang2017knowledge}. Existing KGs often fail to capture the fine-grained details and hierarchical structure of academic papers, resulting in a loss of valuable information and context. Moreover, querying and retrieving relevant knowledge from large-scale academic KGs remains a daunting task, as complex queries often lead to inefficient and inaccurate results \cite{verma2023scholarly}.

To address these limitations, there is a pressing need for more advanced and intelligent approaches to construct and utilize academic KGs. In this paper, we present a novel framework that aims to enhance the effectiveness and efficiency of academic KG construction and application. Our research introduces two innovative methods: the Deep Document Model (DDM) and the KG-enhanced Query Processing (KGQP).

The DDM is designed to provide a comprehensive and fine-grained representation of the hierarchical structure and semantic relationships within academic papers. By employing advanced Natural Language Processing (NLP) techniques, such as text segmentation and named entity recognition, DDM transforms unstructured text data into a structured knowledge representation. This approach enables a more nuanced and context-rich representation of scholarly knowledge, surpassing the limitations of traditional KGs.

On the other hand, KGQP focuses on optimizing complex queries over large-scale academic KGs. By leveraging the structural information and semantic relationships encoded in the KG, KGQP simplifies and optimizes user queries, significantly improving query accuracy and efficiency. Through techniques such as entity recognition, relevance ranking, and graph traversal, KGQP enables users to retrieve the most relevant and informative knowledge from the vast academic KG.

A key innovation of our framework is the integration of KGs with state-of-the-art LLMs. By combining the structured knowledge representation of DDM with the powerful language understanding capabilities of LLMs, our approach achieves effective knowledge utilization and enhances the performance of various downstream tasks, such as question answering, document summarization, and knowledge discovery.

To evaluate the effectiveness of our proposed framework, we conduct extensive experiments on real-world academic datasets. The experimental results demonstrate that our approach significantly outperforms existing methods in terms of both KG construction accuracy and query efficiency. Furthermore, we showcase the practical application of our framework in academic research scenarios, highlighting its potential to revolutionize the way researchers access, analyze, and utilize scholarly knowledge.

The main contributions of this paper can be summarized as follows:

\begin{enumerate}

\item We propose the DDM, a novel approach for representing the hierarchical structure and semantic relationships of academic papers in a fine-grained manner.
\item We introduce the KGQP method, which leverages the KG structure to optimize complex queries and improve retrieval accuracy and efficiency.

\item We integrate KGs with large language models to enhance knowledge utilization and natural language understanding capabilities.

\item We conduct extensive experiments to demonstrate the superiority of our framework over existing methods in terms of KG construction accuracy and query efficiency.

\item We showcase the practical application of our framework in real-world academic research scenarios, highlighting its potential impact on scholarly knowledge management and discovery. Our research helps researchers acquire and utilize academic knowledge more efficiently, and provides a reference for developing more precise and reliable AI solutions.

\end{enumerate}

The rest of this paper is organized as follows. Section 2 reviews the related work on academic KGs and query optimization techniques. Section 3 presents the proposed Deep Document Model and KG-based Query Optimization method in detail. Section 4 describes the experimental setup and results, followed by a discussion of the practical applications and implications of our framework in Section 5. Finally, Section 6 concludes the paper and outlines future research directions.

\section{Related Work}

In the realm of tools for knowledge extraction from unstructured sources, MEL (Metadata Extractor \& Loader) and TNNT (The NLP-NER Toolkit) provide a formidable combination for parsing a vast array of file formats\footnote{\url{https://w3id.org/kgcp/MEL-TNNT}}. MEL excels in extracting both metadata and text, refining the data into JSON objects for further analysis \cite{mendez2021mel}, while TNNT enhances this process with a robust Named Entity Recognition (NER) pipeline, incorporating numerous NLP models to accurately parse unstructured data \cite{seneviratne2021tnnt}. 

The integration of these tools significantly bolsters the process of extracting pertinent information from academic papers, which is essential for the construction of rich academic KGs. Complementing this toolset, the PARSE component\footnote{\url{https://w3id.org/kgcp/PARSE}} of the KGCP pipeline\footnote{\url{https://w3id.org/kgcp/}} leverages web crawling and advanced NLP models to enrich academic semantic knowledge bases, focusing on the domain of computer science\cite{zhang2023askg}.

Document structure analysis has always been a hot topic in research, with previous researchers proposing various methods\cite{palm2019attend} to analyze and represent the logical structure of documents\cite{Mao2003}. This paper reviews some of the important works in this field, highlighting their contributions and the distinctions from the proposed Deep Document Model. In terms of general document analysis, a method has been developed to classify each heading in a document as representing an entity (its identifier), an attribute, or calculated data, particularly in the context of business process documents, underscoring the potential of analyzing the structural components of documents to efficiently extract relevant information \cite{zhu2018texygen}\cite{van2016storyteller}. In table form document analysis research, the focus has been on document structure grammar to facilitate analysis, with significant efforts made to develop frameworks and techniques for analyzing tabular data within documents \cite{vidgen2020learning}. Moreover, automated methods have been proposed for detecting the structure of regulatory documents, managing tagging and text segmentation, aiming to segment documents into units that can be processed as entries in a database, providing a structured way to represent document content. DDM aims to represent the logical structure of documents at a granular level by providing a robust and flexible framework, representing the development of existing methodologies. Unlike previous works that focused on the physical layout, DDM emphasizes the logical organization of document components, facilitating a range of applications including information extraction, document summarization, comparative analysis, semantic search, and KG construction. It also demonstrates its versatility and adaptability, further contributing to the broader field of document structure analysis and knowledge representation.

\section{Methodology}

\subsection{Document Object Model Ontology (DOMO)}

\begin{figure}[ht]
\tiny
\includegraphics[width=\textwidth]{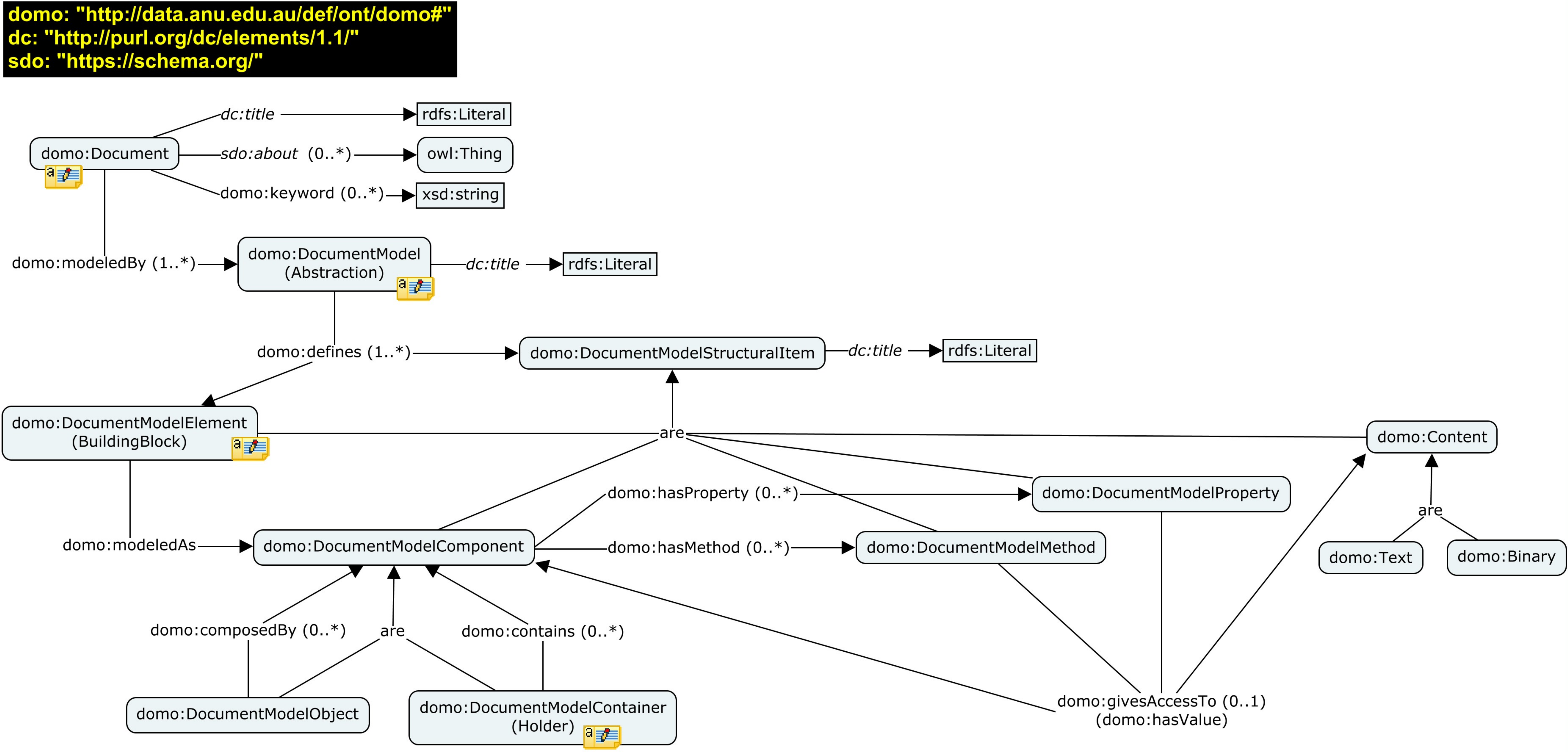}
\caption{Document Object Model Ontology (\textit{DOMO})}\label{DOMO}
\end{figure}

The Document Object Model Ontology (DOMO) is an ontology designed for abstracting, separating, and analyzing the structural components of any document format. It is inspired by the core concepts of the W3C DDM\footnote{\url{https://www.w3.org/DOM/DOMTR}}, which represents the hierarchical structure of web documents (HTML and XML-based) as a tree of interconnected nodes. Similarly, DOMO treats any document (such as academic papers) as a hierarchical structure composed of various logical components (such as title, abstract, sections, paragraphs, sentences, and references). Figure \ref{DOMO} illustrates the main classes and relationships in the DOMO structure.

In this research, we leverage DOMO to develop a systematic approach for extracting and representing the structural and semantic information from academic papers. By applying DOMO-based abstractions, we aim to enhance the granularity and accuracy of the information captured in the ASKG. Moreover, aligning and integrating the DOMO concepts with the existing ASKG ontology structure will enable more sophisticated queries and reasoning over the academic knowledge contained in the ASKG, such as finding papers with similar content at the section, paragraph or sentence level or tracing the flow of ideas and arguments across multiple papers.

By incorporating DOMO concepts into ASKG and the KGCP construction pipeline\footnote{\url{http://w3id.org/kgcp/}}, we seek to create a richer and more expressive representation of academic knowledge that can support advanced applications in research exploration, paper recommendation, and scientific impact assessment.

\subsection{Deep Document Model (DDM)}

Previous ANU Scholarly KGs (ASKGs) have limitations in capturing the structured information of research papers. They can only identify a portion of the sections in a paper, leading to incomplete knowledge. To overcome this limitation, we introduce the DDM.

The DDM aims to transform unstructured text data of academic papers into a structured knowledge representation. In this process, text elements within the document are deeply analyzed using NLP techniques, identifying text elements and their hierarchical relationships. This includes text segmentation, and named entity recognition, to extract meaningful components and their attributes from the document.

\begin{figure}[ht]
\tiny
\includegraphics[width=\textwidth]{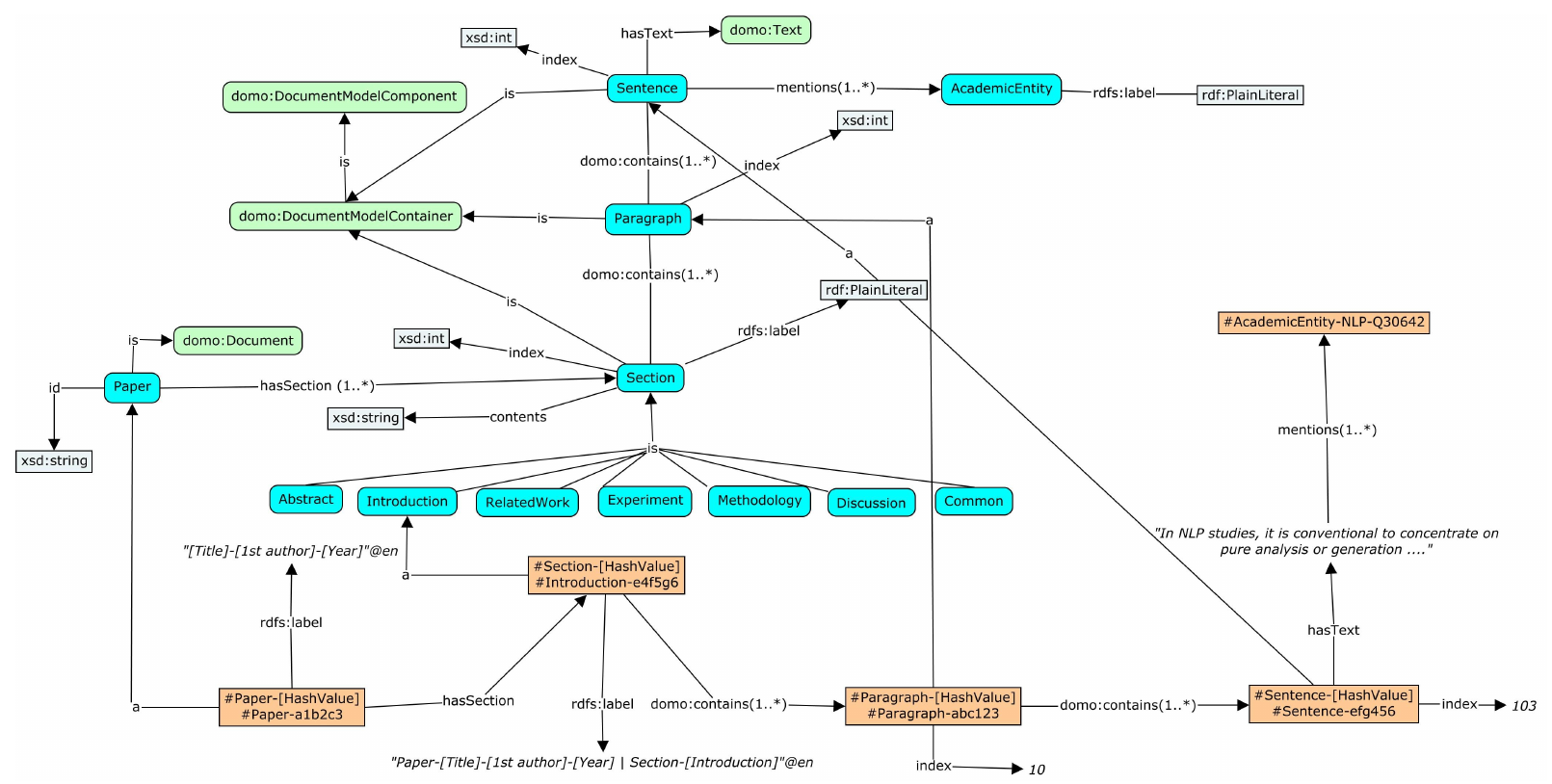}
\caption{Deep Document Model (DDM)}\label{DDM}
\end{figure}

As shown in Figure \ref{DDM}, in building the hierarchical structure of the document, each identified element is considered a node in the hierarchy, with its own type, content, and position attributes within the document. The relationships between nodes are expressed in the form of edges, thereby forming a hierarchical model that reflects the logical flow and structure of the document. To further enhance the accuracy of the model, we also conducted a more detailed decomposition of the document content, breaking it down into smaller sub-components, such as paragraphs, sentences, and entities, and identifying the relationships between these sub-components.

The expanded ASKG using DDM offers the following key advantage: it transcends the traditional role of KGs as mere fact providers. Remarkably, it serves as a comprehensive metadata and context presenter for a collection of research papers. By capturing the logical structure of documents, the new KG offers a more nuanced representation of the information within. Central to this approach is the use of an ontology to create a metadata KG. This research distinguishes itself by presenting an ontology-based KG within a Retrieval-Augmented Generation (RAG) system \cite{lewis2020retrieval}. While other methods rely solely on KG facts, our approach takes a significant step further by harnessing the power of ontologies to enhance the depth and richness of the knowledge representation.

\begin{figure}[htbp]
\tiny
\includegraphics[width=\textwidth]{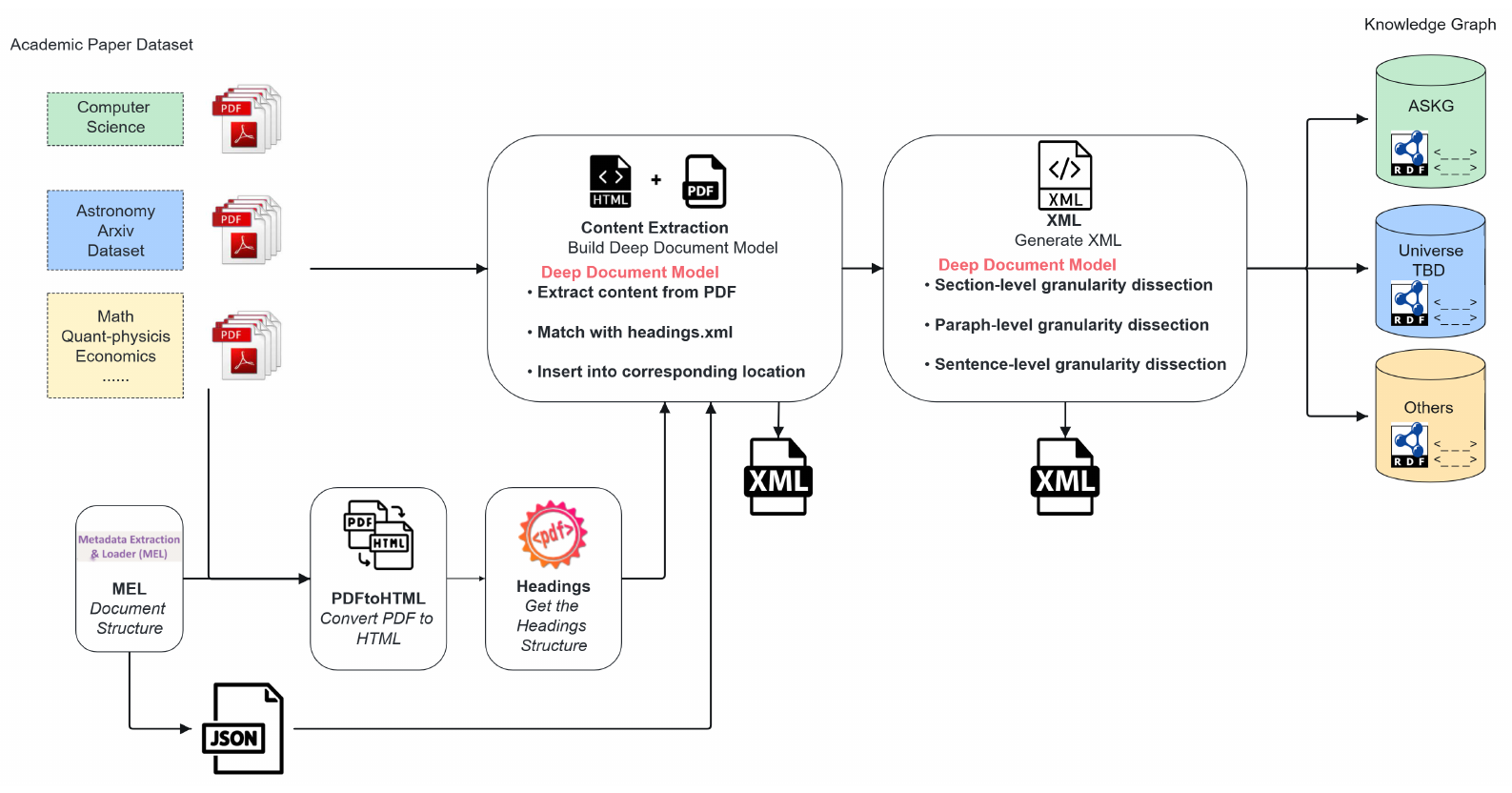}
\caption{Deep Document Model Pipeline}\label{DDMPipeline}
\end{figure}

Figure \ref{DDMPipeline} shows the entire DDM construction process. Initially, the academic document is transitioned from PDF to HTML format using the PDFtoHTML tool, facilitating the extraction of its hierarchical heading structure. This is subsequently integrated with the outcomes derived from the MEL Document Structure and the original PDF document, culminating in the formation of the Deep Document Model. The process commences with the extraction of content from the PDF file, segmenting it into individual sentences. These sentences are then meticulously aligned with the headings obtained from the XML document, ensuring their accurate placement within the appropriate sections. The model functions at various levels of granularity, encompassing sections, paragraphs, and singular sentences. It also identifies any references as specific embedded elements linked to those sentences. Ultimately, the XML document is converted into RDF, thereby enriching the Scholarly KG. This enrichment paves the way for additional applications, notably facilitating interactions with LLMs as explored in this paper.

In the specific construction process of the document model, based on hierarchical decomposition, we developed a model encapsulating the logical structure of the document, integrating relationships and component attributes (such as importance, relevance) into the model. For ease of subsequent processing and analysis, this model is serialized into XML or JSON format. As shown in Listing \ref{code:chunking_sample}, it constructs a document structure containing multiple levels of granularity. In the chunked part, each section is defined by a \texttt{\textless section\textgreater} and uniquely identified by an ID. The document includes titles (using \texttt{\textless heading\textgreater}) and text content (using \texttt{\textless sentence\textgreater}), as well as related reference numbers (using \texttt{\textless reference\textgreater}). Moreover, the code demonstrates how to nest different sections, forming a hierarchical document structure. Additionally, we have utilised a visualization tool to intuitively display the structure and content of the document, facilitating user understanding and exploration.
\clearpage

\begin{lstlisting}[
    language=XML,
    basicstyle=\ttfamily\small,
    label=code:chunking_sample,
    frame=single, 
    rulecolor=\color{black},
    framerule=0.5mm, 
    framesep=3mm,
    breaklines=true,
    caption={Examples of Chunking Result in XML},
    captionpos=b
]
<section>
<section ID="1">
<heading>Introduction</heading>
<sentence>In the 21st century, the importance of developing cutting-edge scientific research is self-evident for every country.</sentence>
<reference>1</reference>
<sentence>Usually, the government research funding agencies receive thousands of research proposals each year, which are reviewed only by expert panels.</sentence>
</section>
<section ID="2">
<heading>Related Work</heading>
<section ID="2.1">
<heading>Computer science in evaluating grant applications</heading>
<sentence>Oztaysi et al.</sentence>
<sentence>
proposed a multi-criteria approach to evaluate research proposals based on interval-valued intuitionistic fuzzy sets.
<reference>2</reference>
</sentence>
<sentence>
Besides the IC score, reviewers also score several other assessment scores, such as "Feasibility Score" or "Significance Score." </sentence>
<sentence>
A Metadata Extractor & Loader (MEL) tool is applied to extract text from PDF research proposals and save it in a JSON file with metadata sets and content.
<reference>20</reference>
</sentence>
<sentence>
By default, all JSON files are stored in CouchDB database based on the proposal index.
<reference>21</reference>
</sentence>
........
\end{lstlisting}

Currently, our primary focus is on academic papers, particularly in the field of computer science. However, we plan to extend DDM to encompass a wider range of disciplines, including astronomy and others. The versatility of our model lies in its ability to adapt to a variety of document structures, enabling future implementations to handle diverse document types ranging from recommendation letters and business documents to PowerPoint presentations. We have already established a use case of ASKG for computer science papers, but it can be applied to any domain by constructing a domain-specific ontology and following our pipeline. In addition to academic papers, we are also applying DDM to other applications, such as the UniverseTBD project\footnote{\url{https://universetbd.org/}} in the field of astronomy.

This study presents an innovative approach in document analysis, utilizing DDM to advance academic research and knowledge discovery. As a testament to the evolving landscape of document analysis, DDM demonstrates significant potential for structuring and interpreting complex documents. Future work will focus on enhancing algorithmic efficiency and integrating advanced NLP techniques, thereby ensuring the model's continuous evolution and relevance in intuitive and insightful data interpretation.

\subsection{KG-enhanced Query Processing (KGQP)}

To address the challenge of ``AI hallucination'' faced by LLMs when dealing with complex questions, particularly those involving intricate fact verification, we have designed and implemented the KG-enhanced Query Processing (KGQP) workflow.\footnote{\url{https://w3id.org/kgcp/KGQP}} This process harnesses the potential advantages of LLMs and incorporates an academic KG aligned to DOMO. KGQP is a pipeline that provides KG-based context to LLMs in question-answer interactions.  The main idea is that ontology-based KGs, such as ASKG, hold structured information (facts and metadata -- such as the DDM) that could provide accurate context to prompt engineering tasks with LLMs to mitigate their ``hallucinations''. In the following sections, we will provide a detailed analysis of the various aspects of this processing flow.

\subsubsection{Graph Matching in the KG}
In a nutshell, KGQP facilitates graph matching between LLM Output Triples (LOT) produced by LLMs and a KG. This process provides a foundation for generating answers via LLMs and is divided into two core tasks:

Firstly, SPARQL queries are used for either ``exact matching'' or ``fuzzy matching'' to extract a set of Candidate Triples from the KG (CTKG). This step finds potential matches within the KG for the LOT derived from user queries.

Secondly, after obtaining these CTKG, interaction with the LLMs is conducted to determine the final KG triples that match with the user-generated LOT. This interactive process not only ensures the precise selection of triples from the CTKG that match with LOT, but it also provides a more accurate and reliable basis for answers generated by LLMs. The final output consists of a set of KG triples that match with the LOT, which will serve as the baseline data for subsequent research and applications.

% \begin{figure}[ht]
% \tiny
% \includegraphics[width=\textwidth]{img/LLM_Interact_Flow_Chart.png}
% \caption{LLMs Interaction Flow Chart}\label{LLM_Interact}
% \end{figure}

\begin{figure}[p]
    \centering
    \includegraphics[width=\textheight,height=\textwidth,keepaspectratio,angle=90,origin=c]{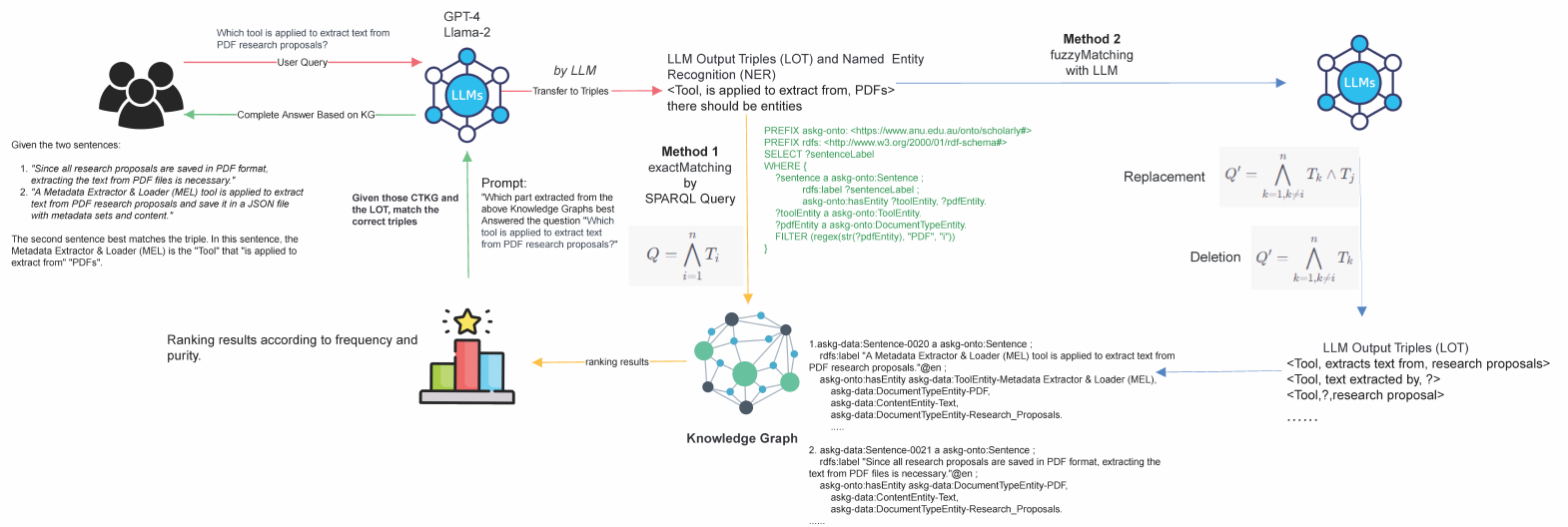}
    \caption{LLMs Interaction Flow Chart}
    \label{LLM_Interact}
\end{figure}

Figure \ref{LLM_Interact} shows the entire triple mapping process. In the first step of our experiment, the user submits a query to the LLM. For this study, we have chosen to utilize LLaMA2, a moderately performing LLM, rather than a state-of-the-art model. The rationale behind this decision is to ensure that the generated responses are primarily influenced by our KG rather than the LLMs’ own training data. By employing a less advanced model, we aim to minimize the potential bias introduced by the LLMs’ pre-existing knowledge and focus on evaluating the effectiveness of our KG in guiding the generation process. Unlike traditional LLMs, which directly process and answer the user's query, in our methodology, the LLMs first transform the user's query into a form of triples, referred to as LOT.

Then, we attempt to perform an exact matching of triples with the KG. In the exact mapping, we use the LOT transformed by the LLM for SPARQL queries. For example, a simple query, ``Which tool is applied to extract text from PDFs?'', can be directly mapped to a triple \( \langle \texttt{Tool}, \texttt{is applied to extract from}, \texttt{PDFs} \rangle \). However, for more complex queries, a single triple often cannot capture the full semantics of the question. In such cases, it may be necessary to construct a compound query consisting of multiple triples.

Consider a complex query $Q$, which can be decomposed into $n$ triples $T_1, T_2, \ldots, T_n$. These triples can be connected by logical conjunctions (``AND'' expressions) to form a SPARQL query:

$$
Q = T_1 \land T_2 \land \ldots \land T_n
$$

or in mathematical notation:

\begin{equation}
Q = \bigwedge_{i=1}^{n} T_i
\end{equation}

Here, we construct a SPARQL query template to transform the logical representation of the query into a SPARQL query statement by LLMs with a predefined prompt model, as follows:
\begin{lstlisting}[
    language=XML,
    basicstyle=\ttfamily\small,
    label=code:sparql_sample,
    frame=single, 
    rulecolor=\color{black},
    framerule=0.5mm, 
    framesep=3mm,
    breaklines=true,
    caption={Example of the SPARQL Query Template},
    captionpos=b
]
PREFIX askg: <https://www.anu.edu.au/onto/scholarly#>
PREFIX rdfs: <http://www.w3.org/2000/01/rdf-schema#>

SELECT ?entity ?label WHERE {
  ?entity a askg-onto:Paragraph
   rdfs:label ?label .
  FILTER (
    {conditions}
  )
}
\end{lstlisting}

The purpose of this SPARQL query is to retrieve paragraph entities from the KG. The query initializes by defining an entity (\texttt{?entity}) and selecting its label (\texttt{?label})\footnote{The textual content of each paragraph is stored as a ``label'' of the KG entity}. It ensures that the selected entities are entries of type \texttt{askg-onto:Paragraph} and their labels are selected. Additional refinement in the filtering conditions can be used to match paragraphs containing any specified entity names. Specifically, this is achieved by converting the paragraph labels to lowercase and checking whether these labels contain the lowercase names of the specified entities. This process involves using string functions for matching and verifying the names within the labels. This method enables the query to identify and retrieve paragraphs linked to the specified entities, which contain relevant information or data concerning user query entities. Ultimately, the query results are utilized to calculate the frequency of specific keywords within each paragraph and rank and filter the paragraphs based on this frequency, thereby identifying the most relevant paragraphs.

By performing SPARQL queries in an automatically constructed KG, results in Listing \ref{code:result_example} can be obtained.

\begin{lstlisting}[
    language=XML,
    basicstyle=\ttfamily\small,
    label=code:result_example,
    frame=single, 
    rulecolor=\color{black},
    framerule=0.5mm, 
    framesep=3mm,
    breaklines=true,
    caption={Example of the SPARQL Query Results},
    captionpos=b
]
askg-data:Excerpt-33387384e82242 a askg-onto:Excerpt ;
    rdfs:label "Paper-[''] | Section-['Introduction'] | Excerpt-[9153]-[9155]"@en ;
    askg-onto:inSentence "for poker and ya core we can find no previously prepared data , so we randomly partition 90 % for training and 10 % for testing"^^xsd:string ;
    askg-onto:mentions askg-data:AcademicEntity-prepared_data ;
    askg-onto:wordIndexFrom "9153"^^xsd:int ;
    askg-onto:wordIndexTo "9155"^^xsd:int .

askg-data:Excerpt-33521e403cb12f a askg-onto:Excerpt ;
    rdfs:label "Paper-[''] | Section-['Introduction'] | Excerpt-[5834]-[5836]"@en ;
    askg-onto:inSentence "computed em representations of pre and entities in he functions that inform the generation of io rules bounded by a maximum length"^^xsd:string ;
    askg-onto:mentions askg-data:AcademicEntity-io_rule_bound ;
    askg-onto:wordIndexFrom "5834"^^xsd:int ;
    askg-onto:wordIndexTo "5836"^^xsd:int .
\end{lstlisting}

The excerpt is generated by the PARSE pipeline, which identifies and extracts academic entities. Further details can be found in Section 4.3. However, the exact matching strategy usually might fail, especially when the number of $T_i$ in the search criteria becomes too large: it becomes prohibited to retrieve sentences or paragraphs containing all $T_i$. In such cases, we employ a``fuzzy matching'' strategy to execute the query with a ``relaxation'' mechanism. \textit{Query Relaxation} can be more precisely described using mathematical and logical symbols.

We define an operation $ \text{Relax}(Q, i, j) $ that takes a composite query $ Q $ and returns a new composite query in which the $ i $-th triple $ T_i $ is replaced by or removed for the $ j $-th triple $ T_j $.

\begin{itemize}
    \item \textbf{Replacement}: If $ T_i $ is replaced by $ T_j $, the new query $ Q' $ becomes:

    \begin{equation}
    Q' = \bigwedge_{k=1, k\neq i}^{n} T_k \land T_j
    \end{equation}

    \item \textbf{Deletion}: If $ T_i $ is removed, the new query $ Q' $ becomes:

    \begin{equation}
    Q' = \bigwedge_{k=1, k\neq i}^{n} T_k
    \end{equation}
\end{itemize}

Based on this operation, we can define a \textit{Query Relaxation} function $ \text{RelaxSet}(Q) $, which returns a set $ W $ containing all possible relaxed queries:

\begin{equation}
W = \text{RelaxSet}(Q) = \{ \text{Relax}(Q, i, j) : i, j \in \{1, \ldots, n\} \text{ and } i \neq j \}
\end{equation}

$ i $ and $ j $ can refer to a library or dictionary of replacement triples, not just the triples present in the original query $ Q $.

In this manner, $ W $ encompasses all possible queries that can be obtained by replacing or deleting one or multiple triples in $ Q $. This offers a systematic way to dynamically adjust the complexity of queries, thereby enhancing the flexibility and effectiveness of KG querying.

In the scenario depicted in Figure \ref{LLM_Interact}, when exact matching fails to yield results, we may resort to employing replacement or deletion strategies to simplify the query conditions. For instance, if the original query triple \texttt{<Tool, extracts text from, research proposals>} cannot be directly matched within the KG, a replacement operation can be executed. This operation might involve substituting certain elements with other entities or relationships. The modified query could then be \texttt{<Tool, text extracted by, ?>}, where the question mark (?) denotes a fuzzy match, which could correspond to any action or entity related to the tool. Moreover, the deletion operation allows us to remove certain entities from the query, such as in \texttt{<Tool, ?, research proposal>}, where the question mark represents the omitted part of the triple that specifies extracting text from research proposals, thereby simplifying the query condition.

This approach allows us to flexibly adjust the complexity of the query and find results within the KG that are highly relevant to the original inquiry through a fuzzy matching strategy. Not only does this process enhance the flexibility of querying, but it also improves the effectiveness of KG searches. Ultimately, by using this method of \textit{Query Relaxation}, we can obtain results akin to those presented in listing \ref{code:result_example}, providing foundational data for further research and applications.

Such a flexible method for handling queries is crucial when dealing with KGs, especially in the face of complex and variable user queries. By appropriate replacement and deletion operations, we can ensure the provision of high-quality results, even when confronted with ambiguous or incomplete data.

\subsubsection{Results Ranking and User Interaction}
To rank the results of SPARQL queries, we employ two metrics: frequency and purity. The frequency formula \( F(E) \) measures the occurrence of a specific Entities \( E \) in the KG. Specifically, frequency can be expressed as:
\begin{equation}
F(E) = \text{Count of } E \text{ in Extracted Triples}
\end{equation}
The purity formula \( P(E) \), on the other hand, focuses on the clarity and contextual relevance of the triple in the related text, and can be defined as:
\begin{equation}
P(E) = \frac{\text{Number of Relevant Entities of } E}{\text{Total Entities of } E \text{ in Extracted Triples}}
\end{equation}
After obtaining the set of CTKGs, we use specific prompts to guide the LLM in analyzing and ranking the returned triple results. For example, a prompt might be: Which part extracted from the above KGs best Answered the question “Which tool is applied to extract text from PDF research proposals?”
\\
Upon processing this prompt, the LLM returned the following results:

\begin{quotation}
\noindent This sentence directly states that the \texttt{``Metadata Extractor \& Loader (MEL)''} tool is used for extracting text from PDF research proposals, which is precisely the information required by the question.
\end{quotation}

This method not only enables the LLM to identify information most relevant to the user query but also enhances the understanding and utilization efficiency of the KG. This LLM-based interaction and result ranking approach provide an effective solution for retrieving information from complex text and KG.

\section{Experiments}
In this section, we conducted experiments to determine the effectiveness of our proposed KG-based information retrieval system. Our system leverages a KG with a specific ontology to ingest a set of documents and uses it to retrieve relevant context for a given user query. The code and results are available at the following persistent web ID: \href{http://w3id.org/kgcp/KGQP}{http://w3id.org/kgcp/KGQP}.

As our baseline, we meticulously employed the most common method of document ingestion and retrieval in a Retrieval-Augmented Generation (RAG) system. This method, known for its reliability, utilizes fixed-size chunking with overlap for ingestion and vectorization for retrieving the most relevant chunks for a given query. We will delve into the exact experimental settings of the baseline in the following subsection, leaving no room for ambiguity.

To compare our KG-based information retrieval system for RAG with the baseline method, we conducted three sets of experiments:

\begin{itemize}
    \item Comparison based on the embedding of retrieved contexts with the given query: In this experiment, we used embedding representation to determine the relevance of the retrieved chunks to the query. Since this evaluation closely aligns with the retrieval method of the baseline, it has an inherent advantage. However, we included this measure to assess how different chunking strategies in the baseline and our system affect the relevance of the retrieved chunks.
    \item End-to-end evaluation: In this experiment, we used a sample of queries and employed the same LLM-based answer generation module to create responses based on the retrieved context. We evaluated the final responses based on various qualities, including relevance, accuracy, completeness, and readability. Two expert evaluators and one compatible LLM evaluator compared the final responses of the two rival systems.
    \item Comparison of the relevance of retrieved chunks regarding their named entities: This measure provided a more explainable relevance metric, focusing on the presence and significance of named entities in the retrieved chunks.
\end{itemize}

We used ten peer-reviewed research papers on KG construction and completion, and Brain-Computer Interfaces for the source documents. Our expertise in this domain facilitated the identification of expert evaluators.

\subsection{Vector-based baseline experiment settings}
%no need for subsubsection as each of them is a paragraph
%\subsubsection{Data Preprocessing}

In the baseline experiment, a simple chunking approach was employed to preprocess the text data. The dataset used in this experiment consists of 10 scientific papers in the field of computer science. These papers were all published by the same two authors, who are also the experts involved in the subsequent human evaluation. The dataset was divided into chunks with a maximum of 100 tokens per chunk and a 5\% overlap ratio between adjacent chunks. On one hand, this ensures that the content length of each chunk obtained by ``Simple Chunking'' is comparable to the average length of Paragraphs\footnote{These paragraph entities were extracted through the DDM pipeline.} in follow experiments. On the other hand, it provides richer contextual information. Appropriate overlapping can make the semantic connections between adjacent chunks closer, thus providing more effective support for subsequent semantic similarity calculations and paragraph selection. This chunking process yielded a total of 618 text chunks.

%\subsubsection{Embedding Vector Generation}

To represent the text chunks and user queries in a vector space, the DistilBERT model \cite{sanh2019distilbert} was utilized to generate embedding vectors. Each text chunk and user query was processed using the DistilBERT model, resulting in a set of embedding vectors that capture the semantic information of the text.

%\subsubsection{Similarity Calculation and Selection}

The cosine similarity between each text chunk and the user query was computed to measure their semantic relevance. Based on the similarity scores, the top N most similar chunks were selected for different experimental conditions, with N varying as 1, 3, 5, 10, and 15. These selected chunks served as the context for the subsequent query answering step.

%as this module is shared between baseline and KG-based, should be independent 
%you have two rival for context providing 
\subsection{Generating answer based on the given context module}

The LLaMA2 model was employed to generate answers to the user queries based solely on the selected context chunks, without utilizing any external data. Five specific questions were chosen for evaluation, covering various aspects related to the provided text:

\begin{enumerate}
\item In terms of KG completion, what are the key challenges in ensuring accuracy and completeness?
\item How do different data mining techniques compare when analyzing complex datasets in computer science research?
\item What role does machine learning play in enhancing the capabilities of brain-computer interactions?
\item How can ontology-based models improve the efficiency of data processing in IoT environments?
\item What are the emerging trends in named entity recognition and how do they impact information extraction?
\end{enumerate}

These questions were specifically selected because they are directly relevant to the provided text but cannot be easily answered by LLMs without the given context.

The generated answers by LLaMA2 for each question were recorded and saved for comparison with the results obtained from the subsequent KG experiment.

\subsection{KG-based solution experiment settings}

\subsubsection{KG Construction}

First, we extracted information from 10 scientific papers using the PARSE\footnote{\url{http://w3id.org/kgcp/PARSE}} and DDM\footnote{\url{http://w3id.org/kgcp/DDM}} methods, and constructed a KG file (.ttl file). The KG detail is shown in table \ref{tab:kg_metrics}.

The resultant KG contains a total of 107 sections and 259 paragraphs from the 10 articles, with an average of 89 words per paragraph. In order to associate the summaries containing academic entity information extracted from PARSE with the paragraphs extracted from DDM, we used the GIST embedding model to generate a new relation: \texttt{askg-onto:Paragraph} has \texttt{askg-onto:Excerpt}. There is a total of 3,847 excerpts in our KG. By setting a similarity threshold of 0.7, 3,697 (96.1\%) of these excerpts were successfully linked to their corresponding paragraphs. There are 21 types of entities relationships and 12 types of entities. The total number of triples is 40,507 and the total number of entities is 26,058. Through this approach, we integrated the results from DDM and PARSE into a complete KG for subsequent research use.

\begin{table}[H]
\centering
\caption{Knowledge Graph Construction Metrics}
\label{tab:kg_metrics}
\begin{tabular}[1.5]{lr}
\toprule
Item & Value \\
\midrule
Number of scientific papers & 10 \\
Total sections & 107 \\
Total paragraphs & 259 \\
Average words per paragraph & 89 \\
Total excerpts in the KG & 3,847 \\
Excerpts linked to paragraphs & 3,697 \\
Percentage of linked excerpts & 96.1\% \\
Number of relationship types & 21 \\
Number of entity types & 12 \\
Number of triples & 40,507 \\
Number of entities & 26,058 \\
\bottomrule
\end{tabular}
\end{table}

The following is an example of matching results:

\begin{lstlisting}[
    language=XML,
    basicstyle=\ttfamily\small,
    label=code:paragraph_result,
    frame=single,
    rulecolor=\color{black},
    framerule=0.5mm,
    framesep=3mm,
    breaklines=true,
    caption={Example of SPARQL Matched Paragraphs},
    captionpos=b
]
askg-data:Paper-0be8d70b82d0d7-Paragraph-b294e6f2db8caf2cffc735e60347a90a a askg-onto:Paragraph ;
    rdfs:label "The content of this specific paragraph."@en ;
    askg-onto:hasExcerpt askg-data:Excerpt-063537230ee499,
        askg-data:Excerpt-08dc3cd7733974,
        askg-data:Excerpt-19c38f90f99868,
\end{lstlisting}

\subsubsection{Entity Recognition and Matching}

We employed LLMs (GPT-4 \cite{achiam2023gpt} and LLaMA) to identify entities in user queries and match these entities with the academic entities in the KG. Through the KG, we can retrieve the paragraph content related to the query. We attempted to use two LLMs, GPT-4 and LLaMA, for entity recognition and matching. After comparison, we found that LLaMA2's performance was relatively weak, and it was not satisfactory in terms of the accuracy and quantity of entity recognition. In contrast, GPT-4 performed exceptionally well, accurately identifying entities in the questions and providing corresponding quantities. Therefore, we ultimately chose the entity recognition results of GPT-4 for subsequent KG matching. GPT-4's powerful language understanding and generation capabilities laid the foundation for achieving precise entity matching.

\clearpage
Below is an example details for one of the paragraphs:

\begin{lstlisting}[
    language=,
    basicstyle=\ttfamily\small,
    label=code:paragraph_details,
    frame=single,
    rulecolor=\color{black},
    framerule=0.5mm,
    framesep=3mm,
    breaklines=true,
    caption={Details of a Matched Paragraph},
    captionpos=b
    float=H
]
Paragraph 1 URI: https://www.anu.edu.au/onto/scholarly/Paper-0e8235511ed563-Paragraph-dd82029abfc6b2957c1f393e79ff1411
Paragraph 1 Text: The most comprehensive and current state-of-the-art tool for content extraction and analysis is Apache Tika, which is a complete and complex Java-based general-purpose system. While MEL core goals resemble the ones of Apache Tika, the main difference and benefit of MEL as compared to Apache Tika is that it is a lightweight Python-based package for the metadata extraction of common file formats aimed to be used in a KGCP.
Keyword Frequency: 5
\end{lstlisting}

%Provide tables to show the characteristics of constructed KG for the specific set of prepares

%Too much unnecessary subsubsection

% \clearpage
\subsubsection{Paragraph Selection}

We adopted the Keyword Frequency Matching method to select the top N paragraphs (N = 1, 3, 5, 10, 15) most relevant to the query. Among the top 10 paragraphs, we used the diversity evaluation method to select the top 5 most diverse paragraphs.

\subsubsection{Query Answering}
%as it is the same as the what discussed in the previous subsection no need for further eplaination
The selected 5 diverse paragraphs were used as context and fed into the LLaMA2 model to generate answers to user queries. LlaMA2 will only consider the provided content as context instead of using its pretrained knowledge.

\subsection{Evaluation}

We employed human evaluation methods in this experiment. In the human evaluation, we invited two experts and Claude \cite{claude2023} as three judges to score the answers generated by the two experiments, evaluating the answers in four dimensions: Relevance, Accuracy, Completeness, and Readability. We designed a 5-point scoring standard to ensure consistency in the evaluation.

Additionally, we extracted all entities from the results obtained by the ``Simple Chunking'' and ``KG Approach'' methods and calculated their overlap and Jaccard distance. We also calculated the embedding distance. In the embedding distance evaluation, we used GIST-Embedding to compare the paragraphs obtained in the vector-based experiment and the KG-based experiment and calculated the semantic distance between them. A smaller distance indicates that the paragraphs selected by the KG experiment are more similar to the results of the baseline.

We recorded the number of article sources for each answer to analyze the context diversity of the two methods. We also calculated the similarity between the answers obtained by different methods and the embeddings of user queries. The results showed that the KG-based solution was almost entirely higher than Simple Chunking, indicating that the KG-based method produced significantly different results compared to the Simple Chunking method, demonstrating superior performance.

Finally, we compared and analyzed the evaluation results of the three judges on different questions and used Cronbach's Alpha to calculate the internal consistency reliability of the scale. The result was 0.867, indicating that the evaluation results have good consistency.

\subsubsection{Human Evaluation Analysis}

Next, we compare the human evaluation scores of the two experiments across four dimensions: Relevance, Accuracy, Completeness, and Readability. The evaluators are both authors of the papers. The results are presented in Table \ref{tab:comparison}.

\begin{table}[h!]
\centering
\begin{tabular}{lcccc}
\toprule
& Relevance & Accuracy & Completeness & Readability \\
\midrule
Evaluator 1 & \\
\quad KG-based & 3.4 & \textbf{3.8} & \textbf{3.4} & 4.0 \\
\quad Vector-based & 3.4 & 3.4 & 2.8 & 4.0 \\
\midrule
Evaluator 2 & \\
\quad KG-based & 3.6 & \textbf{3.6} & \textbf{3.6} & 4.2 \\
\quad Vector-based & 3.6 & 3.4 & 3.2 & 4.2 \\
\midrule
Evaluator 3 & \\
\quad KG-based & \textbf{4.6} & \textbf{4.0} & \textbf{4.0} & \textbf{4.6} \\
\quad Vector-based & 4.2 & 3.8 & 3.2 & 4.0 \\
\midrule
Average & \\
\quad KG-based & \textbf{3.9} & \textbf{3.8} & \textbf{3.7} & \textbf{4.3} \\
\quad Vector-based & 3.7 & 3.5 & 3.1 & 4.1 \\
\bottomrule
\end{tabular}
\caption{Comparison of KG-based and Vector-based Systems on five questions}
\label{tab:comparison}
\end{table}

Table \ref{tab:comparison} presents the average scores given by three evaluators for the KG-based and vector-based systems across four key dimensions: Relevance, Accuracy, Completeness, and Readability. The evaluators, including two experts and the Claude3 AI assistant, assessed the generated answers to five questions using a 5-point scale.

The results show that the KG-based method performs well across all dimensions. In terms of Relevance, the KG-based method achieves an average score of 3.9, indicating that the answers generated by the KG-based method are generally aligned with the given questions. The vector-based method achieves a comparable average score of 3.7 in this dimension. Accuracy is another dimension where the KG-based method demonstrates its capability. The KG-based method obtains an average Accuracy score of 3.8, suggesting that the answers generated by the KG-based method are generally precise and reliable. The vector-based method achieves an average Accuracy score of 3.5. In terms of Completeness, the KG-based method achieves an average score of 3.7, indicating that the answers generated by the KG-based method are relatively comprehensive and provide a reasonably thorough response to the given questions. The vector-based method obtains an average Completeness score of 3.1. Readability is a dimension where both methods perform well, with the KG-based method achieving an average score of 4.3 and the vector-based method scoring 4.1. This suggests that the answers generated by both methods are generally well-structured, coherent, and easy to understand.

Looking at the individual evaluator scores, we can observe that the KG-based method maintains a consistent performance across all dimensions for each evaluator. Evaluator 3, in particular, shows a higher preference for the KG-based method, with relatively higher scores in Relevance, Completeness, and Readability. The average scores in the last row of the table provide an overview of the KG-based method's performance. Across all dimensions, the KG-based method maintains reasonably high average scores, indicating its ability to generate answers that are relevant, accurate, complete, and readable.

These findings suggest that the KG-based method's approach of leveraging the meta-data and structure of the document to provide context can be effective in generating answers to given questions. By utilizing the information captured in the knowledge graph, the KG-based method is able to provide answers that are generally aligned with the questions, precise, reasonably comprehensive, and easy to understand. The results presented here provide a promising foundation for exploring the benefits of integrating knowledge graphs into question answering systems.

In conclusion, the evaluation results highlight the potential of the KG-based method in generating answers that are relevant, accurate, complete, and readable. By leveraging the meta-data and structure of the document through the knowledge graph, the KG-based method offers a promising approach to providing context-aware answers to given questions. Further research and development in this area may lead to enhancements in the performance and usability of question answering systems.

\subsubsection{Embedding Distance Analysis}
In the embedding distance evaluation, we used GIST-Embedding to encode the chunks retrieved in the vector-based baseline experiment and the DDM paragraphs in the KG-based experiment. We then calculated the cosine similarity between the embeddings to measure the semantic distance. A higher cosine similarity score (i.e., a smaller semantic distance) indicates that the retrieved paragraphs selected by the KG-based experiment are more semantically similar to the results of the vector-based baseline. The results are shown in Table \ref{tab:embedding_distance}.

\newcolumntype{C}[1]{>{\centering\arraybackslash}p{#1}}

\begin{table}[htbp]
\centering
\caption{Internal Embedding Distance Comparison}
\label{tab:embedding_distance}
\begin{tabular}{C{4cm} C{4cm}}
\toprule
\textbf{Question} & \textbf{Embedding Distance} \\
\midrule
Q1 & 0.7188 \\
Q2 & 0.6852 \\
Q3 & 0.7808 \\
Q4 & 0.9121 \\
Q5 & 0.6847 \\
\bottomrule
\end{tabular}
\end{table}

The results in Table \ref{tab:embedding_distance} show the cosine similarity between the embeddings of chunks vs. paragraphs retrieved by the vector-based baseline and the KG-based experiment for each question. The cosine similarity scores range from 0.6847 to 0.9121, with an average of 0.7563.
These scores indicate that there are differences between the chunks and paragraphs selected by the two methods. Q4 has the highest cosine similarity of 0.9121, suggesting that the chunks and paragraphs retrieved by both methods for this question are highly similar. However, the other questions (Q1, Q2, Q3, and Q5) have lower cosine similarity scores, with an average of 0.7174. This implies that the KG-based method retrieves paragraphs that are semantically different from those chunks selected by the vector-based baseline.

The differences between the retrieved chunks and paragraphs can be attributed to the additional contextual information provided by the knowledge graph in the KG-based experiment. The KG-based method likely considers the relationships and semantic connections between entities in the knowledge graph, which may lead to the selection of paragraphs that are more relevant to the specific context of the question. These findings suggest that integrating a KG into the retrieval process can lead to the selection of different and potentially more relevant ``self-contained textual content'' (paragraphs) compared to a purely vector-based approach (chunks).

\subsubsection{Entity Analysis of KG-based and Vector-based Answers}

Lastly, we examine the entity-related metrics of the generated answers from both experiments. Table \ref{tab:entity_comparison} shows the Overlap Entity Ratio and Jaccard Distance between the two sets of answers for each question.

\newcolumntype{C}[1]{>{\centering\arraybackslash}p{#1}}

\begin{table}[htbp]
\centering
\caption{Entity Analysis between KG-based and Vector-based Answers}
\label{tab:entity_comparison}
\begin{tabular}{C{4cm} C{4cm} C{4cm}}
\toprule
\textbf{Question} & \textbf{Overlap Entity Ratio} & \textbf{Jaccard Distance} \\
\midrule
Q1 & 0.0833 & 0.9545 \\
Q2 & 0.0625 & 0.9630 \\
Q3 & 0.1304 & 0.8966 \\
Q4 & 0.1875 & 0.8750 \\
Q5 & 0.3478 & 0.7391 \\
\bottomrule
\end{tabular}
\end{table}

Table \ref{tab:entity_comparison} presents the entity-related metrics comparing the answers generated by the KG-based experiment and the vector-based baseline. The Overlap Entity Ratio measures the proportion of entities that are common to both sets of answers, while the Jaccard Distance quantifies the dissimilarity between the two entity sets.

The results show relatively low Overlap Entity Ratios, ranging from 0.0625 to 0.3478, indicating that the KG-based experiment generates answers containing a significantly different set of entities compared to the baseline. Correspondingly, the Jaccard Distances are high, with values above 0.7 for all questions, further confirming the dissimilarity in entity composition between the two approaches.
These findings suggest that by leveraging the entity information stored in the knowledge graph, the KG-based experiment can identify and include a more diverse and relevant set of entities in the generated answers. The rich semantic connections captured in the KG enable the system to better understand the relationships between entities and concepts, allowing it to retrieve and incorporate more pertinent information in the responses. The inclusion of a wider range of relevant entities in the KG-based answers can provide users with additional informative details and context, potentially leading to better satisfaction of their information needs.

In summary, the entity-related metrics highlight the differences in entity composition between the answers generated by the KG-based experiment and the vector-based baseline. The low overlap ratios and high Jaccard distances suggest that the KG-based approach can produce answers with a more diverse and potentially relevant set of entities by leveraging the structured information captured in the KG. However, additional evaluation of the quality and usefulness of the answers is required to draw definitive conclusions about the superiority of the KG-based method in question-answering tasks.

In this study, we propose a KG-based question answering system that integrates semantic retrieval to improve the accuracy and completeness of answers. Although the current experiments were conducted on a relatively small dataset, the results demonstrate the potential of KGs in question-answering tasks. We observed that by integrating the KG, the proposed method could retrieve more relevant context paragraphs and generate answers containing a more diverse and relevant set of entities. These findings suggest that leveraging the structured information in KGs can effectively improve the performance of question-answering systems. Therefore, we believe that the proposed approach has good scalability and is capable of operating and being applied to larger-scale datasets.

Our approach shares some commonalities with the latest research advances, such as Azure AI Search's hybrid retrieval + semantic ranking method \cite{azuresearch2023}. Both methods employ semantic retrieval and KG techniques to enhance the performance of question answering systems. Specifically, our baseline utilizes embedding vector-based semantic retrieval, which is similar to Azure AI Search's vector retrieval. Our KG experiment leverages the KG for entity recognition and matching, and combines keyword frequency matching and diversity assessment to select relevant paragraphs (extracted via the DDM pipeline), which is conceptually similar to Azure AI Search's hybrid retrieval approach. This demonstrates that our research direction aligns with the current state-of-the-art in the field.

Drawing from the experience of Azure AI Search, we can further optimize and improve our method in future work. For instance, we can consider introducing a more sophisticated hybrid retrieval strategy, adjusting the chunking strategy, incorporating semantic ranking methods, expanding evaluation metrics, and increasing the scale of experiments. Through these enhancements, we aim to further boost the performance of our question answering system and provide more convincing experimental support for the application of KGs in question answering tasks. We believe that with continuous research and optimization of methods, KGs will play an increasingly crucial role in the domain of question answering systems, offering more intelligent and efficient means for people to acquire knowledge and information.

In conclusion, this study demonstrates the significant potential of integrating KGs into question answering systems. The experimental results provide strong evidence that leveraging the structured knowledge captured in KGs can substantially enhance the relevance, accuracy, completeness, and diversity of generated answers. The proposed approach offers a promising direction for developing more intelligent and effective question answering systems that can better understand and satisfy user information needs.

\section{Conclusion}
In this paper, we presented a novel approach to enhance the construction and application of academic KGs. Our research aimed to address the limitations of traditional KGs by introducing two innovative methods: the DDM and the KGQP. The DDM provides a comprehensive and fine-grained structured representation of document content, enabling a more nuanced and context-rich representation of scholarly knowledge. KGQP leverages the KG to optimize and simplify complex queries, significantly improving query accuracy and efficiency.

A key contribution of our work is the integration of DDM with existing document management systems and KG frameworks, which significantly enhances the structuring and analysis of academic papers. By employing advanced NLP techniques, our method effectively transforms unstructured text into structured knowledge, facilitating deeper insights and understanding. Moreover, the combination of KGs with LLMs achieves effective knowledge utilization and enhances natural language understanding capabilities.

Experimental results demonstrate that the integration of KGs substantially improves the performance of question answering systems. The rich background knowledge provided by the KG enables the generation of more accurate and complete answers. Through entity recognition, paragraph selection, and other key steps, our method identifies relevant information and provides diverse contexts pertinent to the query.

Our methodology, combining the strengths of document decomposition, KG integration, and enhanced query processing, marks a significant step forward in the realm of academic paper analysis and knowledge discovery. It paves the way for more accurate, reliable, and efficient AI-powered solutions, with applications extending across various sectors. The implications of our research are far-reaching, offering new perspectives and methodologies for the fields of KG and natural language processing.

In conclusion, our research contributes substantially to the fields of document structure analysis, knowledge representation, and NLP, offering a robust and flexible framework that enhances the capabilities of LLMs. By addressing critical challenges such as AI hallucination, we aim to foster the development of more precise, trustworthy, and efficient AI-powered tools for academic and industry applications. As we continue to push the boundaries of KG research, we remain committed to driving innovation and transforming the way researchers access, analyze, and apply academic knowledge.

\section{Discussion}

Through the baseline and the KG experiment, we have demonstrated the effectiveness of incorporating KGs in enhancing the performance of question answering systems. The evaluation results show that leveraging the structured information from the KG can improve the relevance, accuracy, and completeness of the generated answers compared to the baseline approach.

These promising findings motivate us to further explore the potential of KGs in various aspects of the question answering task. In this section, we propose and discuss several extension directions to optimize and enhance the KG-based question answering approach, including diversity extension, structural distance, RAG method with KG information fusion, KG reasoning, and explainable question answering. These extensions aim to fully utilize the rich information embedded in the KG and provide more comprehensive, reliable, and interpretable answers to user queries.

\subsection{Diversity Extension}
In addition to selecting the most diverse paragraphs, we propose considering more dimensions to measure the diversity of nodes in the KG, such as authors, research institutions, research topics, publication years, and citation relationships. To achieve this, a diversity scoring function can be designed to comprehensively consider the diversity of different dimensions and calculate an overall diversity score for each node. For example, for a given node, we can consider the number and influence of its authors, the reputation of the research institution, the breadth of research topics, the span of publication years, and the diversity of citation relationships to calculate its overall diversity score. When selecting paragraphs, not only the relevance to the query should be considered, but also the diversity of nodes, striving to cover different research perspectives and backgrounds as much as possible. For instance, for a query about ``KG applications'', we can select relevant paragraphs from different authors, institutions, topics, and years to provide comprehensive and multi-faceted information.

\subsection{KG Structural Distance}
We suggest leveraging the structural information of the KG to compute the structural distance between nodes. Classical graph distance algorithms, such as shortest path distance and PageRank distance, can be employed. For example, in a scientific article KG, we can calculate the shortest path distance between two research topic nodes to measure their relevance in the research field. Additionally, we can compute the PageRank score of an author node to assess its importance and influence in the academic network. When representing each node in the embedding space, not only the semantic information of the node itself should be considered, but also its structural information in the KG. Methods such as TransE \cite{bordes2013translating}, TransR \cite{lin2015learning}, RotatE \cite{sun2019rotate}, and Graph Neural Networks (GNNs) \cite{scarselli2008graph} can be utilized to encode the structural features of nodes into embedding vectors. For instance, using the TransE model, we can embed author nodes and research topic nodes into the same vector space and capture the structural information by leveraging the translation relationship between them. During the paragraph selection process, maximizing the structural distance of selected nodes in the KG should be considered to ensure that the retrieved information comes from different regions of the graph, thereby enhancing the diversity and comprehensiveness of the information.

\subsection{RAG Method with KG Information Fusion}
The KG can be regarded as an external knowledge base, and the idea of RAG can be applied to dynamically retrieve relevant information from the KG during the answer generation process. We can retrieve more fine-grained knowledge, such as entities, relations, and attributes. Specifically, during the process of LLaMA2 generating answers, based on the currently generated content, relevant nodes and relations in the KG can be retrieved in real-time and used as additional contextual information to guide the subsequent generation process. For example, when generating an answer about ``KG applications'', if the currently generated content mentions ``KG completion'', we can instantly retrieve entities (e.g., missing relations, entities) and relations (e.g., ``head entity-relation-tail entity'' triples) related to ``KG completion'' from the KG and incorporate this information into the subsequent answer generation. By doing so, we can utilize the existing information in the KG to infer the missing parts and generate more complete and accurate answers. Furthermore, we can retrieve other entities and relations related to ``KG completion'', such as commonly used KG embedding models (e.g., TransE, ComplEx) and evaluation metrics (e.g., Mean Reciprocal Rank, Hits@N), to enrich the content of the answer and provide more comprehensive information. By dynamically fusing KG information, the RAG method can significantly enhance the quality and reliability of answer generation, ensuring that the generated answers are consistent with the KG while covering a broader range of relevant knowledge.

\subsection{KG Reasoning and Question Answering}
By leveraging the reasoning capability of the KG, higher-level question-answering tasks, such as multi-hop QA and logical reasoning QA, can be realized. For example, given a complex query like ``Researcher A published a paper on topic C at institution B, which was cited by researcher D. What are the research interests of researcher D?'', we can perform multi-hop traversal and reasoning in the KG. First, we can find the node representing researcher A, then follow the ``publish'' relation to find the paper node, next follow the ``cite'' relation to find the node of researcher D who cited the paper, and finally follow the ``research interest'' relation to find the research interest node of researcher D. Through this multi-hop reasoning, we can generate an accurate answer.

\subsection{Explainable Question Answering Based on KG}
The structured information of the KG can be utilized to generate explainable question-answering results, providing not only the answer but also the reasoning process and evidence behind it. The relevant nodes and relations in the KG can serve as the basis for explanation, generating natural language reasoning chains to help users understand the source and logic of the answer. For example, for the aforementioned complex query, we can generate the following reasoning chain: ``Researcher A published a paper on topic C at institution B → The paper was cited by researcher D → Researcher D's research interests are E'', clearly showing the reasoning process from the question to the answer. Experiments can evaluate the quality of explainable question answering and user satisfaction and trust in the explainability. We can invite users to assess the clarity, relevance, and credibility of the reasoning chains and collect feedback to further improve the explainability.

\section{Future Work}

Our study has introduced an innovative approach and framework for LLMs in academic text processing, yet there remain numerous areas for further exploration and enhancement. The following are the directions for our future work:

\subsubsection{Dynamic Update and Learning of KG}
We propose exploring how to dynamically update and extend the KG during the question-answering process to adapt to new scientific articles and knowledge. Language models can be utilized to extract key information, such as entities, relations, and attributes, from new articles and dynamically integrate them into the existing KG. For example, when a new scientific article is published, we can use named entity recognition and relation extraction techniques to extract new researchers, institutions, topics, and other entities from the article, as well as the relationships between them, such as ``author-publish-paper'' and ``paper-belong to-topic''. Then, we can add these new entities and relations to the existing KG, expanding its coverage and depth. Experiments can assess the impact of the dynamically updated KG on question-answering performance, as well as the expandability and flexibility of the KG. We can compare the performance of static KGs and dynamically updated KGs in answering new questions and evaluate the KG's ability to adapt to new knowledge.

\subsubsection{Multimodal KG and Question Answering}
Non-textual information in scientific articles, such as figures and formulas, can also be incorporated into the construction and question-answering process of the KG. Figures and formulas can be transformed into structured representations, such as figure KGs and mathematical formula trees, and fused with the textual KG. For example, for a scientific article containing a line chart, we can extract entities (e.g., variables, trends) and relations (e.g., correlations between variables) from the chart to construct a figure KG. Additionally, we can link the figure KG with the textual KG, associating the variables in the chart with the concepts mentioned in the text. During the question-answering process, we can utilize the multimodal KG to answer complex questions involving figures and formulas, such as ``According to the chart, what kind of relationship is presented between variable A and variable B?'' or ``What does symbol Y represent in formula X?''. Experiments can evaluate the impact of the multimodal KG on question-answering performance, especially for problems that require synthesizing information from text, figures, and formulas. We can compare the accuracy and completeness of answering such questions using only the textual KG versus using the multimodal KG.

\textbf{Enrichment} Currently, DDM primarily focuses on textual content. We aim to extend its capabilities to include further content structural decomposition of tables, mathematical formulas, equations, and charts. This will enable the DDM to identify and parse non-textual elements in documents, such as data charts and illustrations, providing additional dimensions for a comprehensive understanding of academic papers.

\textbf{Automate Pipeline} We will work on automating the PARSE and DDM pipeline processes. This automation will significantly enhance our ability to handle large-scale document collections, providing robust support for big data analysis.

Through these future directions, we aim to further enhance the capabilities of LLMs in handling academic texts, especially in the areas of entity recognition, data enrichment, automation, and data interoperability. These improvements will significantly increase the efficiency and depth of academic research, bringing new breakthroughs in the fields of knowledge discovery and academic analysis.

\footnotesize
\bibliographystyle{splncs04}
\bibliography{ref.bib}

\end{document}